\documentclass[12pt]{article}
\usepackage{amssymb,amsmath,amsthm,amsfonts,amscd}
\usepackage{graphicx}
\newcommand\be{\begin{equation}}
\newcommand\ee{\end{equation}}

\newcommand{\fatalpha}{{\bf \alpha \kern -0.44em \alpha}}
\newcommand{\fatsigma}{{\bf \sigma \kern -0.54em \sigma}}
\newcommand{\tpchi}{{\bf \chi \kern -0.35em \chi}}
\newcommand{\llambda}{{\bf \lambda \kern -0.45em \lambda}}

\bibliography{plain}
 \title{\bf  Comparison of Qubit and Qutrit Like Entangled Squeezed and Coherent States of Light}
 \vspace{20mm}
\author{G. Najarbashi \thanks{E-mail: Najarbashi@uma.ac.ir} ,
  S. Mirzaei \thanks{E-mail: SMirzaei@uma.ac.ir}
\\
{\small Department of Physics, University of Mohaghegh Ardabili, P.O. Box 179, Ardabil, Iran.} \\
 \\
}\pagebreak
\begin{document}
\maketitle \vspace{0mm}

\maketitle \vspace{0mm}
\begin{abstract}
Squeezed state of light is one of the important subjects in quantum optics which is generated by optical nonlinear interactions. In this paper, we especially focus on qubit like entangled squeezed states (ESS's) generated by beam splitters, phase-shifter and cross Kerr nonlinearity. Moreover the Wigner function of two-mode qubit and qutrit like ESS are investigated. We will show that the distances of peaks of Wigner functions for two-mode ESS are entanglement sensitive and can be a witness for entanglement. Like the
qubit cases, monogamy inequality is fulfilled for qutrit like  ESS.
These trends are  compared with those obtained
for qubit and qutrit like  entangled coherent states (ECS).
\\
 {\bf PACs Index: 03.65.Ud}
\end{abstract}

\section{Introduction}
Squeezed states are a general class of minimum-uncertainty states, for which the
noise in one quadrature is reduced compared with a coherent state. The squeezed state of the electromagnetic field can be generated in many optical processes \cite{Slusher,Shelby,Wu}. The first two experiments used third-order nonlinear optical media.
Nonlinear effects such as the Kerr effect can be used to control a light signal by means of another light signal, and
thus play an important role in optical data processing. This nonlinear effect exists in all materials, even isotropic ones, like glass or fused silica. On the other hand, the measurement of the properties of a light field without disturbing it is a necessary requirement of many optical quantum communication and computation protocols which can be achieved by means of the optical Kerr medium with large third-order nonlinearity \cite{Imoto}.
Quantum non-demolition detectors can detect a single photon without destroying it in an absorption process.  This detection relies on the phase shift produced by transmitting a single photon through a nonlinear Kerr medium. Such detectors have recently been implemented in the laboratory \cite{Yamamoto}. The optical Kerr effect also allows the propagation of ultrashort soliton-type pulses without spreading or optical bistability \cite{Deng,Hamedi}. In Ref. \cite{Vivek} the authors have produced large cross-phase shifts of 0.3 mrad per photon with a fast response time using rubidium atoms restricted to a hollow-core photonic bandgap fibre, which shows the largest nonlinear phase shift induced in a single pass through a room temperature medium.
\par
Squeezed state finds a wide range of applications in quantum information processing \cite{van}. A superposition of odd photon number states for quantum information networks has been generated by photon subtraction from a squeezed vacuum state produced by a continuous wave optical parametric amplifier \cite{Neergaard}. In Ref. \cite{Lu}, orthogonal Bell states with entangled squeezed vacuum states have been constructed and a scheme for teleportation a superposition of squeezed states based on the Bell state measurement have been presented. An analysis of squeezed single photon states as a resource for teleportation of coherent state qubits has been investigated in Ref. \cite{Ralph}. Ref. \cite{Kaushik} has shown that non-Gaussian entangled states are good resources for quantum information processing protocols, such as, quantum teleportation. The problem of generating ESS's has been discussed in \cite{Son,Kuang,Zhang,Tipsmark,Wang}. The new physical interpretation of the generalized two-mode squeezing operator has been studied in \cite{Gao} which is useful to design  optical devices for generating various squeezed states of light.
Quantum storage of squeezed state light through an electromagnetically induced transparency ( EIT) medium
 has been experimentally realized in hot atoms and magneto-optical traps in which the atomic quantum memory was verified by measuring the quantum noise of the retrieved states \cite{Harris,Wal,Appel,Honda,Gao1}.
\par
Coherent states, originally introduced by Schrodinger in 1926 \cite{Schrodinger}.
However, multi-mode quantum states of radiation fields which present the opportunity to
be a resource in quantum information theory have been the subject of considerable interest in the experimental and
theoretical literature \cite{Cochrane,Oliveira,Kim,Milburn,Munro,wang3,wang5,wang1,Vogel1,Salimi,Enk,wang2,wang4,najarbashi,Sheng}.
Entanglement concentration for W-type entangled coherent states have been investigated in Ref.\cite{Liu}. Generation of multipartite ECS's and entanglement of multipartite states constructed by linearly independent coherent states have been investigated in \cite{Barry,Enk1}. In Ref. \cite{najarbashi1} the  production and entanglement properties of the generalized balanced N-mode Glauber coherent states of the form
\be\label{mainstate}
|\Psi^{(d)}_{N}\rangle_{c}=\frac{1}{\sqrt{M^{(d)}_N}}\sum_{i=0}^{d-1}f_i|\underbrace{\alpha_i\rangle\cdots|\alpha_i\rangle}_{\mathrm{N \ modes}},
\ee
have been stated which is a general form of  the balanced two-mode ECS $|\Psi^{(2)}\rangle_{c}=\frac{1}{\sqrt{M_{c}^{(2)}}}(|\alpha\rangle|\alpha\rangle+f|\beta\rangle|\beta\rangle)$ and $|\Psi^{(3)}\rangle_{c}=\frac{1}{\sqrt{M_{c}^{(3)}}}(|\alpha\rangle|\alpha\rangle+f_1|\beta\rangle|\beta\rangle+f_2|\gamma\rangle|\gamma\rangle)$. Assuming that the coherent states are linearly independent, these states recast in two qubit and qutrit form respectively. Then the entanglement of these states has been evaluated by concurrence measure. In Ref. \cite{najarbashi2}, the noise effect  on  two modes  qubit and qutrit like ECS's has been studied.
\par
In 1932, Wigner introduced a distribution function in mechanics that permitted a description of mechanical phenomena in a phase space \cite{wigner1,wigner2}. Wigner functions have been especially used for describing the quadratures of the electrical field with coherent and squeezed states or single photon states \cite{Breitenbach,Lvovsky,Agarwal1}. The Bell inequality based on a generalized quasi probability Wigner function  and its violation for single photon entangled states and two-mode squeezed vacuum states have been investigated  in Ref. \cite{Jaksch}. The negativity of the Wigner function \cite{Banerji}  is a reason why the Wigner function
can not be regarded as a real probability distribution but it is a quasi-probability distribution function.
This character  is a good indication of the possibility of the occurrence of nonclassical properties of  quantum states.
The Wigner function of two-mode qubit and qutrit like ECS's have been investigated in \cite{najarbashi3}.
\par
In this paper we will consider two-mode qubit like ESS $|\Psi^{(2)}\rangle_{s}=\frac{1}{\sqrt{M_s^{(2)}}}(|\xi\xi\rangle+f|\eta\eta\rangle)$, in which two squeezed states $|\xi\rangle$ and $|\eta\rangle$ are in general nonorthogonal and  span a two dimensional qubit like Hilbert space $\{|0\rangle,|1\rangle\}$. Therefore, two-mode squeezed state $|\Psi^{(2)}\rangle_{s}$ can be recast in two qubit form. Moreover as an example we introduce
a method for producing qubit like ESS using kerr medium and beam splitters \cite{Zhang}. The same argument can be formulated for other two-mode squeezed states such as $|\Psi^{(3)}\rangle_{s}=\frac{1}{\sqrt{M_s^{(3)}}}(|\xi\xi\rangle+f_1|\eta\eta\rangle+f_2|\tau\tau\rangle)$, in which the set $\{|\xi\rangle,|\eta\rangle,|\tau\rangle\}$ are linearly independent and  span the three dimensional qutrit like Hilbert space $\{|0\rangle,|1\rangle,|2\rangle\}$, implying that $|\Psi^{(3)}\rangle_{s}$ can be recast in two qutrit form.
The entanglement of the state $|\Psi^{(2)}\rangle_{s}$  can be
calculated by concurrence measure  introduced by Wootters \cite {wooters1,wooters2} and in the same manner, its  generalized version \cite{Akhtarshenas} can be used to measure the entanglement of $|\Psi^{(3)}\rangle_{s}$.
Moreover we investigate the  Wigner quasi-probability distribution function for qubit and qutrit like ESS. For both qubit like ESS and ECS the distance of peaks in Wigner  function can be an evidence for amount of entanglement. While for qutrit like ESS this argument seems to be rather pale.  On the other hand, the monogamy inequality  for multi-qutrit like ESS  and ECS are discussed and it is shown that there  exist qutrit like states violating monogamy inequality.
\par
The outline of this paper is as follows: In section 2 we investigate the entanglement of two-mode qubit like ESS and compare it with ECS. Moreover the Wigner quasi-probability distribution function for qubit like ESS is studied in this section.
 The behavior of Wigner  function and monogamy inequality for multi-qutrit like ESS  are discussed in section 3.
 Our conclusions are summarized in section 4.
\section{Two-mode Qubit like ESS's}
A rather more exotic set of states of the electromagnetic field are the squeezed
states. The squeezed state of light is two-photon coherent state that photons will be created or
destroyed in pairs. They may be generated through the action of a squeeze operator defined as
\be\label{squ}
\hat{S}(\xi)=\exp[\frac{1}{2}(\xi^*\hat{a}^2-\xi\hat{a}^{\dag 2})],
\ee
where $\xi=r_1e^{i\theta_1}$, with $0\leq r_1<\infty$, $0\leq\theta_1\leq2\pi$ and $r_1$ is known as the squeeze parameter. We note that the squeeze operator is unitary and $\hat{S}^{\dag}(\xi)=\hat{S}^{-1}(\xi)=\hat{S}(-\xi)$. The operator $\hat{S}(\xi)$ is a kind of two-photon generalization of the displacement operator used to define the usual coherent states of a single-mode field. Acting squeeze operator on vacuum states would create some sort of two photon coherent states:
\be
|\xi\rangle=\hat{S}(\xi)|0\rangle,
\ee
namely squeezed states. One can write the squeezed states in terms of Fock states $|n\rangle$ as
\be
|\xi\rangle=\frac{1}{\sqrt{\cosh r_1}}\sum_{n=0}^{\infty}(-e^{i\theta_1}\tanh r_1)^n\frac{\sqrt{(2n)!}}{2^n n!}|2n\rangle.
\ee
The overlap of two squeezed state reads
\be\label{dot}
\langle\xi|\eta\rangle=\frac{1}{\sqrt{\cosh r_1\cosh r_2(1-e^{-i\Delta\theta}\tanh r_1\tanh r_2)}},
\ee
in which $\xi=r_1e^{i\theta_1}$, $\eta=r_2e^{i\theta_2}$ and $\Delta\theta=\theta_2-\theta_1$.
\subsection{Entanglement of Two-Mode Qubit like ESS}
\begin{figure}[ht]
\centerline{\includegraphics[width=14cm]{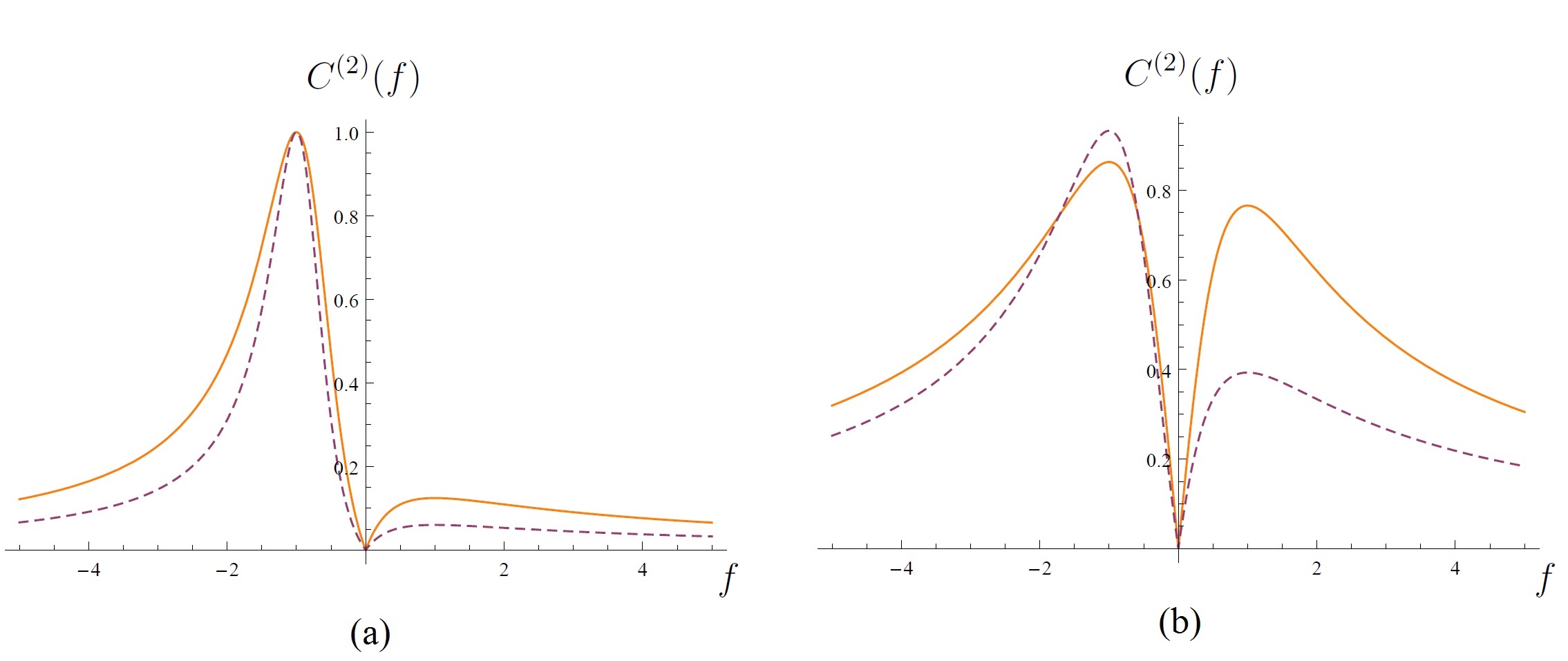}}
\caption{\small {(Color online) Concurrence of $|\Psi^{(2)}\rangle_{s}$ (dashed line) and $|\Psi^{(2)}\rangle_{c}$ (full line) as a function of $f$ for $r_1=1.5$: (a) $r_2=2$ and $\Delta\theta=0$ and (b) $r_2=0.5$ and $\Delta\theta=1.68\pi$ .} \label{1} }
\end{figure}

\begin{figure}[ht]
\centerline{\includegraphics[width=10cm]{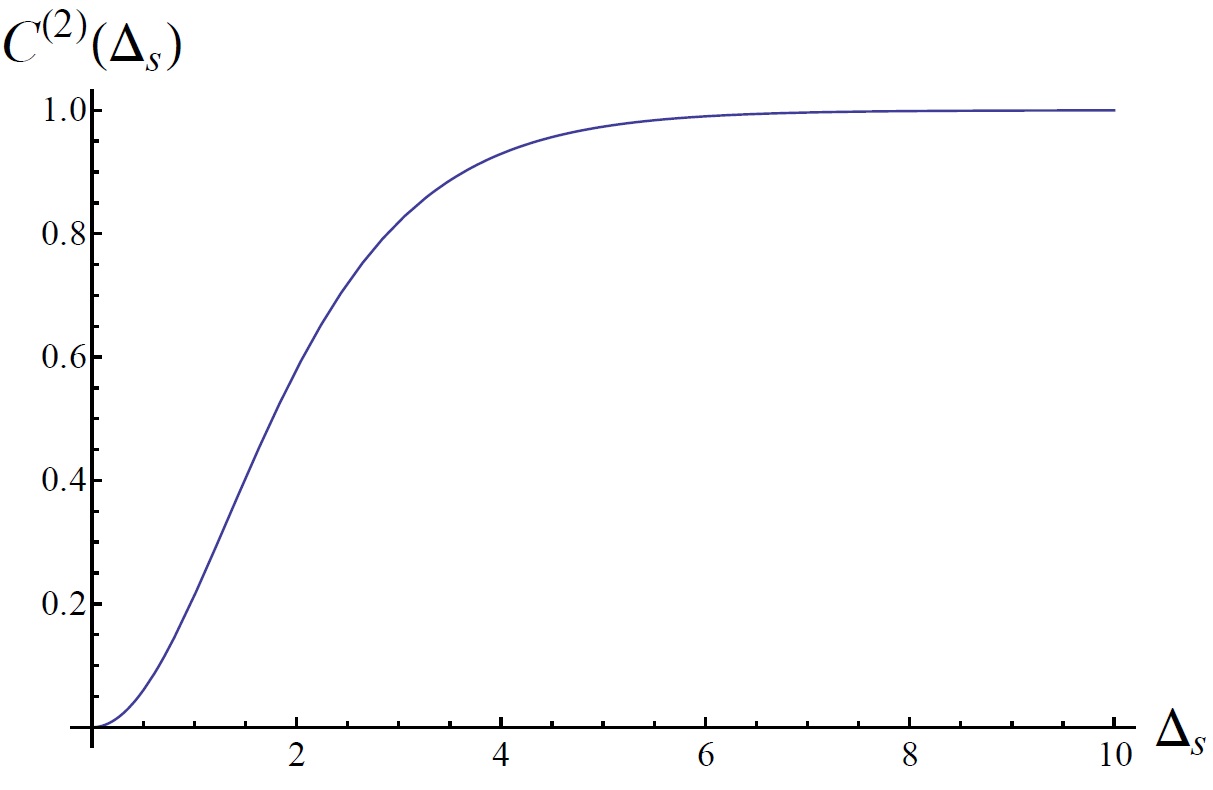}}
\caption{\small {(Color online) Concurrence of $|\Psi^{(2)}\rangle_{s}$ as a function of $\Delta_s=\xi-\eta$ .} \label{12}}
\end{figure}

Let us consider two-mode qubit like ESS as
\be\label{ss}
|\Psi^{(2)}\rangle_{s}=\frac{1}{\sqrt{M_s^{(2)}}}(|\xi\xi\rangle+f|\eta\eta\rangle),
\ee
where $f$, $\xi$ and $\eta$ are generally complex numbers and $M_s^{(2)}$ is a normalization factor, i.e.
\be
M_s^{(2)}=1+|f|^2+2Re(f p^2),
\ee
in which $p=\langle\xi|\eta\rangle$. We used the superscript (2) for qubit-like
states to distinguish it from that of qutrit like states in the next section. Moreover, subscript $s$ is referred to squeezed states to distinguish it from coherent states. Two nonorthogonal squeezed states $|\xi\rangle$ and $|\eta\rangle$ are assumed to be linearly independent and span a two-dimensional subspace of the Hilbert space $\{|0\rangle,|1\rangle\}$. The state $|\Psi^{(2)}\rangle_{s}$ is in general an entangled state. To show this avowal we use a measure of entanglement called concurrence which is introduced by Wooters \cite{wooters1,wooters2}. For any two qubit pure state in the form $|\psi\rangle=a_{00}|00\rangle+a_{01}|01\rangle+a_{10}|10\rangle+a_{11}|11\rangle$, the concurrence is defined as $C=2|a_{00}a_{11}-a_{01}a_{10}|$. By defining the orthonormal basis as
\be
|0\rangle=|\xi\rangle,~~|1\rangle=\frac{1}{\sqrt{1-|p|^2}}(|\eta\rangle-p|\xi\rangle),
\ee
the state $|\Psi^{(2)}\rangle_{s}$ is reduced to the state $|\psi\rangle$. Therefore the concurrence is obtained as
\be\label{res}
C^{(2)}=\frac{2|f|(1-\frac{1}{\cosh r_1\cosh r_2\sqrt{1+\tanh^2r_1\tanh^2r_2-2\cos\Delta\theta\tanh r_1\tanh r_2}})}{1+f^2+\frac{2f(1-\cos\Delta\theta\tanh r_1\tanh r_2)}{\cosh r_1\cosh r_2((1-\cos\Delta\theta\tanh r_1\tanh r_2)^2+\sin^2\Delta\theta\tanh^2r_1\tanh^2r_2)}},
\ee
where for simplicity we assume that $f$ to be a real number. The behavior of concurrence as a function of $f$ is shown in figure \ref{1}. For the sake of  comparison,   the  concurrence of two-mode qubit like ECS,
\be\label{ECS1}
|\Psi^{(2)}\rangle_{c}=\frac{1}{\sqrt{M_c^{(2)}}}(|\alpha\alpha\rangle+f|\beta\beta\rangle),
\ee
 is also plotted in figure \ref{1} (for further details see \cite{najarbashi1}).
A comparison between full and dashed line in figure \ref{1} shows that for $f>0$ the concurrence of squeezed state is less than the concurrence of coherent state. On the other hand for $f<0$, there are $r_1$, $r_2$ and $\Delta\theta$ for which the cross over occurs. Moreover figure shows that only for $f=0$ the state $|\Psi^{(2)}\rangle_{s}$ is separable which is confirmed by Eq.(\ref{res}).
If we assume that all parameters are real and $f=1$, then the concurrence (\ref{res}) can be rewritten as
\be
C^{(2)}(\Delta_s)=\frac{1-\mathrm{sech}[\Delta_s]}{1+\mathrm{sech}[\Delta_s]},
\ee
where $\Delta_s=\xi-\eta$. This equation shows that the concurrence is a monotone function of $\Delta_s$ (see figure \ref{12}). If $\Delta_s\rightarrow\infty$, the concurrence tends to its maximum value ($C^{(2)}_{max}=1$), while for
small separation (i.e. $\Delta_s\rightarrow0$) the concurrence tends to zero which means the state $|\Psi^{(2)}\rangle_{s}$ is separable.
\par
Let us introduce a scheme to generate such ESS's. To this end, consider
the interaction Hamiltonian for a cross Kerr nonlinearity as $\hat{H}=\hbar k \hat{n}_a \hat{n}_b$, where $\hat{n}_a$ and $\hat{n}_b$ are photon number operators of mode $a$ and mode $b$, respectively and $k$ is proportional to the third-order nonlinear susceptibility $\chi^{(3)}$.
The time evolution operator is
\be
\hat{U}(\tau)=e^{-i\tau\hat{n}_a\hat{n}_b},
\ee
where $\tau=kt=k(l/v)$, in which $l$ is the length of Kerr medium (KM) and $v$ is the velocity of light in the Kerr medium. If we assume that mode $a$ is initially in a squeezed vacuum state $|\xi\rangle_{a}$, and choose $\tau=\frac{\pi}{2}$, then we  obtain the superpositions of squeezed vacuum states $|\varphi\rangle\sim|\xi\rangle_a\pm|-\xi\rangle_a$.
\par
The method proposed here, involves cross-Kerr interactions. But, in general, the strengths of these interactions are far too weak to bring about the required phase shifts. The techniques of EIT, a quantum interference phenomenon that arises when coherent optical fields couple to the states of a material quantum system, have been used to  overcome this undesirable  problem. Strictly speaking, the squeezing of the delayed output state in an EIT medium has not been retained  the same as that of input state \cite{Yokoi}, and the squeezing property of the retrieved state after a storage time has decreased \cite{Appel,Honda}. To qualify the efficiency of the squeezing propagation or storage in an EIT system, experimental studies have discovered an approach to repress the noise of output delay light via decreasing the detection frequency or operating the system at off one or two-photon resonance \cite{Cai,Hetet,Xiao,Lukin,Fleischhauer}.
\par
\begin{figure}[ht]
\centerline{\includegraphics[width=10cm]{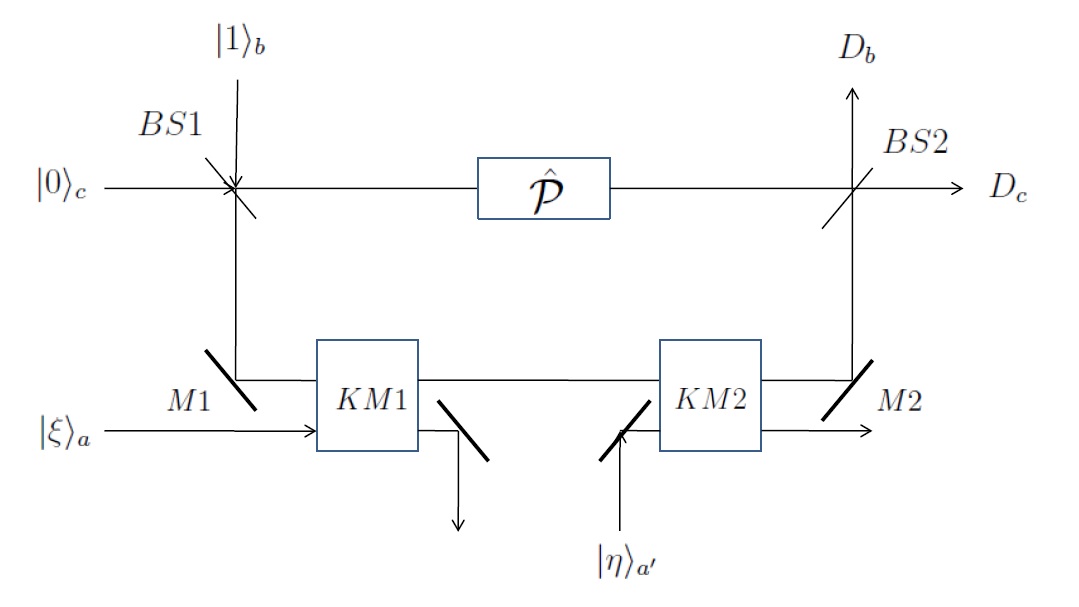}}
\caption{\small {(Color online) Experimental set up for generation ESS.} \label{set}}
\end{figure}
\begin{figure}[ht]
\centerline{\includegraphics[width=10cm]{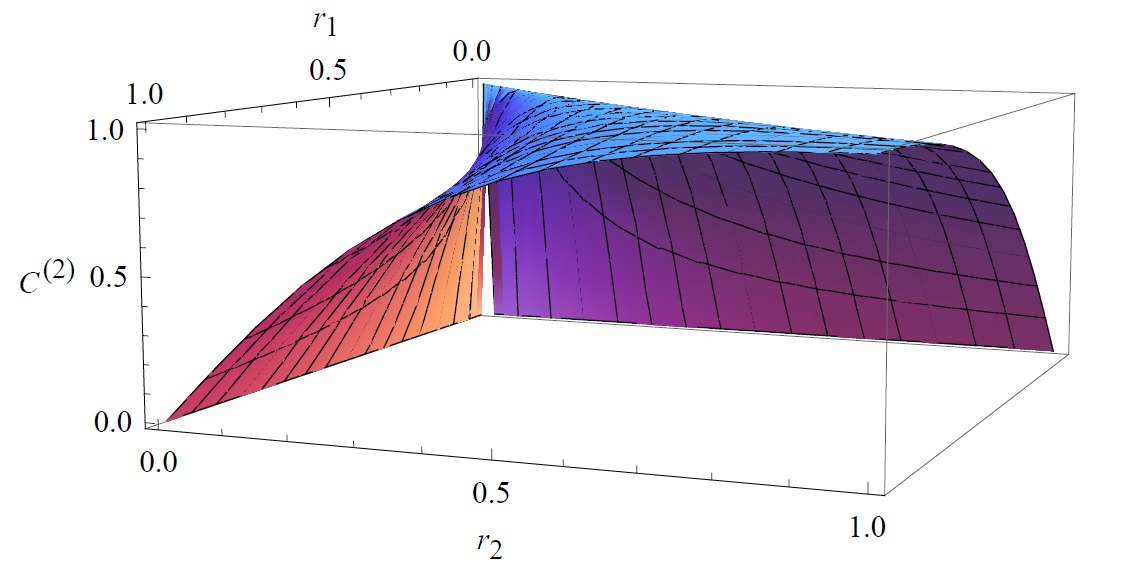}}
\caption{\small {(Color online) Concurrence of $|\psi^{(2)}\rangle_{s}$ as a function of $r_{1}$ and $r_2$.} \label{k}}
\end{figure}

Now, assume that the  state $|1\rangle_b|0\rangle_c$ is inserted to $BS1$, then the output state becomes $\frac{1}{\sqrt{2}}(|10\rangle_{bc}+i|01\rangle_{bc})$ (see figure \ref{set}).
On the other hand, the Kerr medium together with phase shifter $\mathcal{\hat{P}}=e^{i\hat{N}\theta}$,   transform the  state $\frac{1}{\sqrt{2}}|\xi\rangle_a(|10\rangle_{bc}+i|01\rangle_{bc})$ to
\be
|\psi_1\rangle=\frac{1}{\sqrt{2}}(|\xi e^{-2i\tau}\rangle_a|10\rangle_{bc}+ie^{i\theta}|\xi\rangle_a|01\rangle_{bc}),
\ee
where we used the fact that
\be
\hat{U}(\tau)|n\rangle_{a}|\xi\rangle_{b}=|n\rangle_{a}|\xi e^{2i n\tau}\rangle_{b}.
\ee
Now let mode $b$ interacts with mode $a'$ (in a squeezed vacuum state $|\eta\rangle_{a'}$), via Kerr medium $KM2$, then the state $|\psi_1\rangle$ after the Kerr medium $KM2$ and $BS2$ reads
\be
\begin{array}{l}
|\psi_2\rangle=\frac{1}{2}(|\xi e^{-2i\tau}\rangle_a|\eta e^{-2i\tau'}\rangle_{a'}-e^{i\theta}|\xi \rangle_a|\eta \rangle_{a'})|10\rangle_{bc}\\
~~~~~~+\frac{i}{2}(|\xi e^{-2i\tau}\rangle_a|\eta e^{-2i\tau'}\rangle_{a'}+e^{i\theta}|\xi \rangle_a|\eta \rangle_{a'})|01\rangle_{bc}.
\end{array}
\ee
Now if one of the detectors $D_c$ or $D_b$ fires then, up to a normalization factor, we have the ESS's
\be\label{6}
\begin{array}{l}
(|\xi e^{-2i\tau}\rangle_a|\eta e^{-2i\tau'}\rangle_{a'}-e^{i\theta}|\xi \rangle_a|\eta \rangle_{a'}),
\end{array}
\ee
or
\be\label{6}
\begin{array}{l}
i(|\xi e^{-2i\tau}\rangle_a|\eta e^{-2i\tau'}\rangle_{a'}+e^{i\theta}|\xi \rangle_a|\eta \rangle_{a'}).
\end{array}
\ee
Choosing different parameters $\tau,\tau'$ and $\theta$, one may obtain the various  entangled states \cite{Zhang}.
For example,  taking $\tau=\tau'=\frac{\pi}{2}$ and $\theta=0$ we have $|\psi^{(2)}\rangle_s\sim|-\xi\rangle_a|-\eta\rangle_{a'}-|\xi\rangle_a|\eta \rangle_{a'}$ which is  an entangled state with concurrence $C^{(2)}\neq 0$. To see this, we consider
 two nonorthogonal squeezed states $\{|\xi\rangle,|-\xi\rangle\}$ and $\{|\eta\rangle,|-\eta\rangle\}$, which are  linearly independent and span a two-dimensional space  $\{|0\rangle,|1\rangle\}$ defined as
\be\label{base}
\begin{array}{l}
|0\rangle=|\xi\rangle,\quad |1\rangle=\frac{|-\xi\rangle-p_1|\xi\rangle}{N_1}\quad \text{for system 1},\\
|0\rangle=|\eta\rangle,\quad |1\rangle=\frac{|-\eta\rangle-p_2|\eta\rangle}{N_2}\quad \text{for system 2},
\end{array}
\ee
where
\be
\begin{array}{l}
p_1=\langle\xi|-\xi\rangle, \quad N_1=\sqrt{1-p_1^2},\\
p_2=\langle\eta|-\eta\rangle, \quad N_2=\sqrt{1-p_2^2}.
\end{array}
\ee
For simplicity, we again assumed that $\xi$ and $\eta$ are real parameters.
Altogether, we obtain the entangled state
\be
|\psi^{(2)}\rangle_s=\frac{1}{\sqrt{2(1-p_1p_2)}}\{(p_1p_2-1)|00\rangle+N_2p_1|01\rangle+N_1p_2|10\rangle+N_1N_2|11\rangle\},
\ee
whose concurrence reads
\be
C^{(2)}=\frac{\sqrt{(1-p_1^2)(1-p_2^2)}}{1-p_1p_2}.
\ee
The behavior of concurrence as a function of $r_1$ and $r_2$ is shown in figure \ref{k}.
Figure \ref{k} shows that for $r_1=r_2\neq0$ (i.e. $p_1=p_2$) the state $|\psi^{(2)}\rangle_{s}$ is maximally entangled state (i.e. $C^{(2)}_{max}=1$).
\par
\subsection{Wigner Function for Qubit like ESS}
\begin{figure}[ht]
\centerline{\includegraphics[width=14cm]{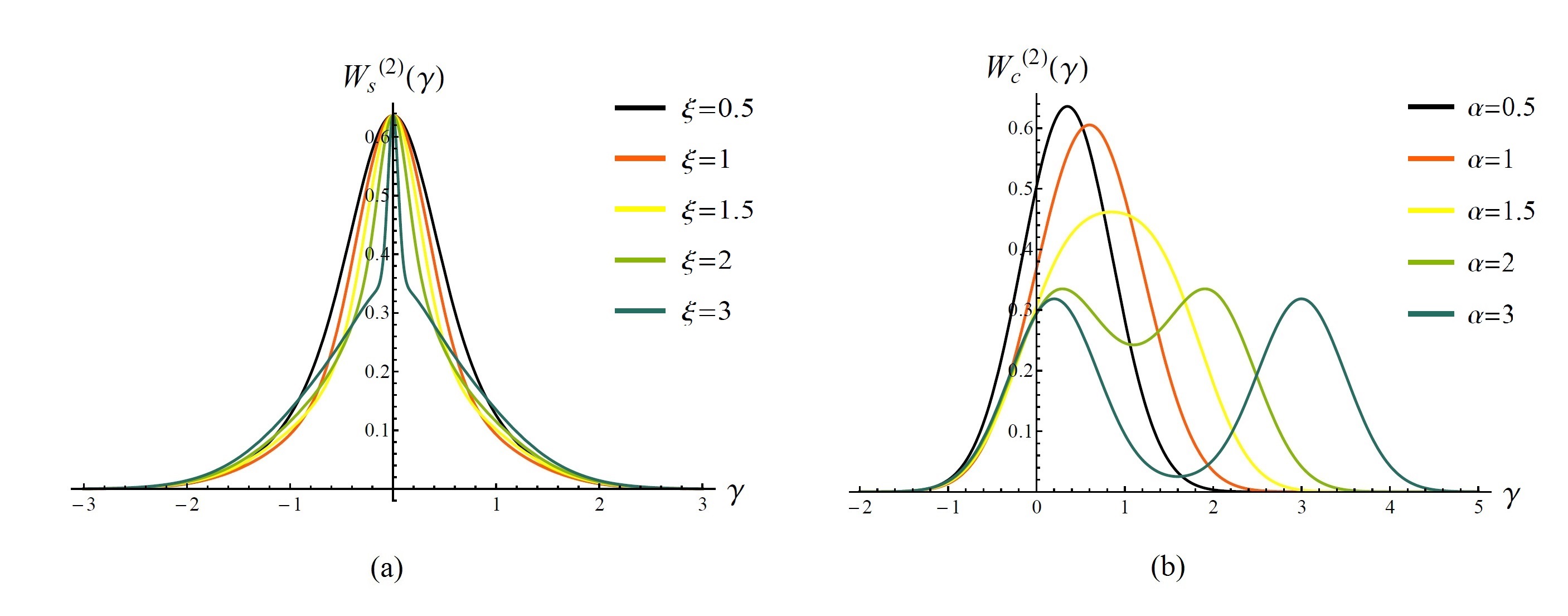}}
\caption{\small {(Color online) Wigner distribution as a function of $\gamma$ with $f=1$ (a) $W_s^{(2)}(\gamma)$ for various $\xi$ and fixed $\eta=0.5$, (b) $W_c^{(2)}(\gamma)$ for various  $\alpha$ and fixed $\beta=0.5$ .} \label{4} }
\end{figure}

\begin{figure}[ht]
\centerline{\includegraphics[width=14cm]{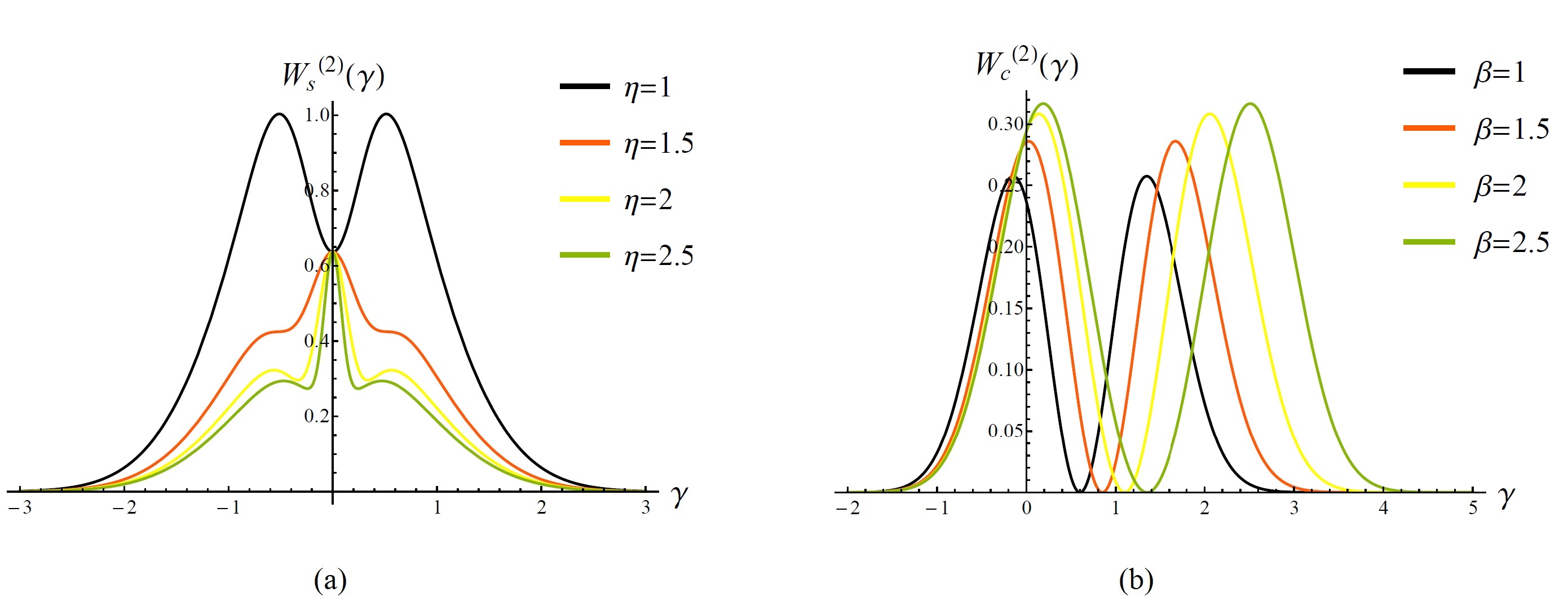}}
\caption{\small {(Color online) Wigner distribution as a function of $\gamma$ with $f=-1$ (a) $W_s^{(2)}(\gamma)$ for given $\eta$, $\xi=0.2$, (b) $W_c^{(2)}(\gamma)$ for given $\beta$, $\alpha=0.2$.} \label{5} }
\end{figure}

One of the important quasi-probability distribution over phase
space is the Wigner function defined as \cite{Book}
\be\label{wigner}
W(\gamma)=\frac{1}{\pi^2}\int d^2\lambda C_W(\lambda)e^{\lambda^*\gamma-\lambda\gamma^*},
\ee
where $C_W(\lambda)=Tr(\rho D(\lambda))$ in which $D(\lambda)=\exp(\alpha \hat{a}^{\dag}-\alpha^{*}\hat{a})$ is displacement operator and $\rho$ is the reduced density matrix  obtained by partially tracing out second mode. Here we investigate the behavior of one-mode Wigner function for qubit like state (\ref{ss}). Wigner characteristic function reads
\be
C^{(2)}_W(\lambda)=\frac{1}{M_s^{(2)}}\{\langle\xi|D(\lambda)|\xi\rangle+f^2\langle\eta|D(\lambda)|\eta\rangle+f p(\langle\xi|D(\lambda)|\eta\rangle+\langle\eta|D(\lambda)|\xi\rangle)\}.
\ee
To calculate the Wigner characteristic function, let us  use the fact that commuting displacement operator $D(\lambda)$ with
squeezing operator $S(\xi)$ yields
\be
S(\xi)D(\lambda)=D(\lambda')S(\xi),
\ee
where $\lambda'=\mu\lambda-\nu\lambda^{*}, \mu=\cosh r_1$ and $\nu=e^{i\theta_1}\sinh r_1.$
By substituting $C^{(2)}_W(\lambda)$ in Eq.(\ref{wigner}) the Wigner function can be obtained as
\be
\begin{array}{l}
W^{(2)}(\gamma)=\frac{1}{M^{(2)}}\{\frac{2}{\pi}(e^{(\frac{-\gamma^2(\mu+\nu)^2}{2})}+f^2e^{(\frac{-\gamma^2(\mu'+\nu')^2}{2})})\\
~~~~+\frac{f p}{\pi\sqrt{2R(\mu\mu'-\nu\nu')(A-B-D)}}\exp[\frac{\gamma^2}{A-B-D}(-1+\frac{(B-D)^2}{2R(A-B-D)})]\\
~~~~+\frac{f p}{\pi\sqrt{2R'(\mu\mu'-\nu\nu')(A'-B'-D')}}\exp[\frac{\gamma^2}{A'-B'-D'}(-1+\frac{(B'-D')^2}{2R'(A'-B'-D')})]
\},
\end{array}
\ee
where
\be
\begin{array}{l}
A=\frac{1}{2}(\mu'^2+\nu'^2-\frac{2\mu'\nu'(\nu'\mu-\mu'\nu)}{\mu\mu'-\nu\nu'}),\quad A'=\frac{1}{2}(\mu^2+\nu^2-\frac{2\mu\nu(\nu\mu'-\mu\nu')}{\mu\mu'-\nu\nu'}),\\
B=\frac{1}{2}(\mu'\nu'-\frac{\mu'^2(\nu'\mu-\mu'\nu)}{\mu\mu'-\nu\nu'}),\quad
B'=\frac{1}{2}(\mu\nu-\frac{\mu^2(\nu\mu'-\mu\nu')}{\mu\mu'-\nu\nu'}),\\
D=\frac{1}{2}(\mu'\nu'-\frac{\nu'^2(\nu'\mu-\mu'\nu)}{\mu\mu'-\nu\nu'}),\quad D'=\frac{1}{2}(\mu\nu-\frac{\nu^2(\nu\mu'-\mu\nu')}{\mu\mu'-\nu\nu'}),\\
R=\frac{1}{2}(A+B+D+\frac{(B-D)^2}{A-B-D}),\quad R'=\frac{1}{2}(A'+B'+D'+\frac{(B'-D')^2}{A'-B'-D'}),
\end{array}
\ee
in which $\mu'=\cosh r_2$ and $ \nu'=e^{i\theta_2}\sinh r_2$.
To explore the behavior of Wigner  function  in more detail, it is convenient
to take  all parameters to be  real except for $\lambda$,  i.e.
 \be
 \begin{array}{l}
 \mu=\cosh\xi,\quad\nu=\sinh\xi,\\
 \mu'=\cosh\eta,\quad\nu'=\sinh\eta.
 \end{array}
\ee
Now we consider two special cases $f=\pm1$. For the case $f=1$ we plot Wigner distribution as a function of $\gamma$ for fixed $\eta=0.5$ but different  $\xi$ (see figure \ref{4} (a)).
Specifically, we wish to compare the results with those obtained for ECS \cite{najarbashi3}.
 By increasing the difference of $\xi$ and $\eta$ the width of squeezed Wigner function $W_s^{(2)}(\gamma)$ decreases, while
the   peaks of $W_c^{(2)}(\gamma)$ for ECS (\ref{ECS1}) recede by increasing the difference of $\alpha$ and $\beta$.
Furthermore,  in contrast to ESS, for the  ECS the height of peaks changes  manifestly .
The behavior of $W_s^{(2)}(\gamma)$ and $W_c^{(2)}(\gamma)$ are depicted in figures \ref{5} for the case $f=-1$.
 Evidently, the peaks of $W_s^{(2)}(\gamma)$ approach to each other by increasing the difference of $\Delta_s=\xi-\eta$. This result is different for $W_c^{(2)}(\gamma)$, that is by increasing the $\Delta_c=\alpha-\beta$, the peaks of $W_c^{(2)}(\gamma)$ recede from each other.
\par
\begin{figure}[ht]
\centerline{\includegraphics[width=10cm]{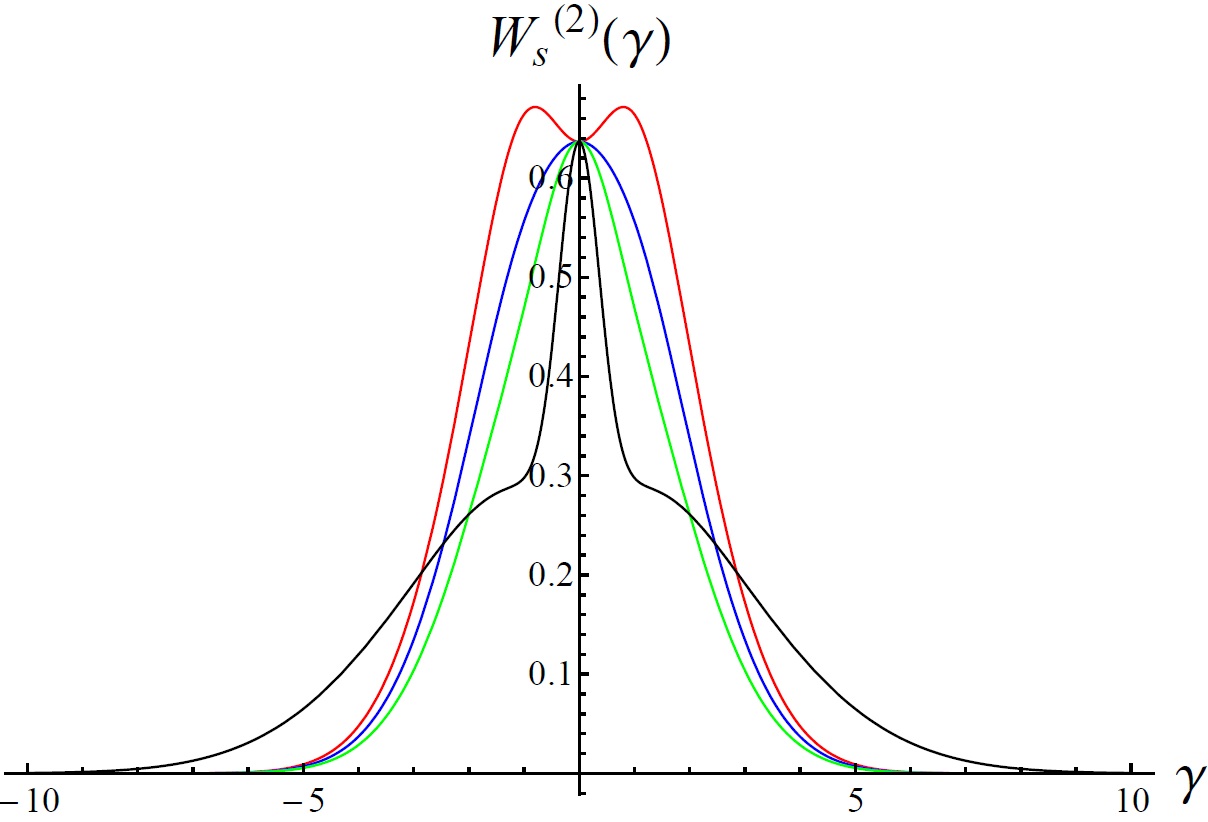}}
\caption{\small {(Color online) $W^{(2)}_s(\gamma)$ as a function of $\gamma$ for given: $\xi=\eta=0.5$ (red), $\xi=0.5$, $\eta=1$ (blue), $\xi=1$, $\eta=2$ (green), $\xi=1$, $\eta=3$ (black).} \label{7} }
\end{figure}

Another example is the qubit like ESS
$|\psi^{(2)}\rangle_s\sim|-\xi\rangle_a|-\eta\rangle_{a'}-|\xi\rangle_a|\eta \rangle_{a'}$), generated by the scheme proposed in figure \ref{set},  whose  Wigner function is given by
\be
\begin{array}{l}
W^{(2)}_s(\xi,\eta,\gamma)=\{\frac{e^{-\frac{1}{2} \gamma ^2 e^{2 \xi }} \left(\cosh (2 \eta ) \left(e^{\gamma ^2 \sinh (2 \xi )}+1\right)-2 \sqrt{\cosh (2 \eta )} \sqrt{\text{sech}(2 \xi )} e^{\frac{1}{4} \gamma ^2 \left(e^{4 \xi }-7\right) \text{sech}(2 \xi )}\right)}{\pi  \left(\cosh (2 \eta )-\text{sech}(2 \xi ) \sqrt{\cosh (2 \eta ) \cosh (2 \xi )}\right)}\}
.
\end{array}
\ee
Figure \ref{7} shows the behavior of the above Wigner function, as a function of $\gamma$, for given $\xi$ and $\eta$.
\section{Qutrit like ESS's}
\begin{figure}[ht]
\centerline{\includegraphics[width=10cm]{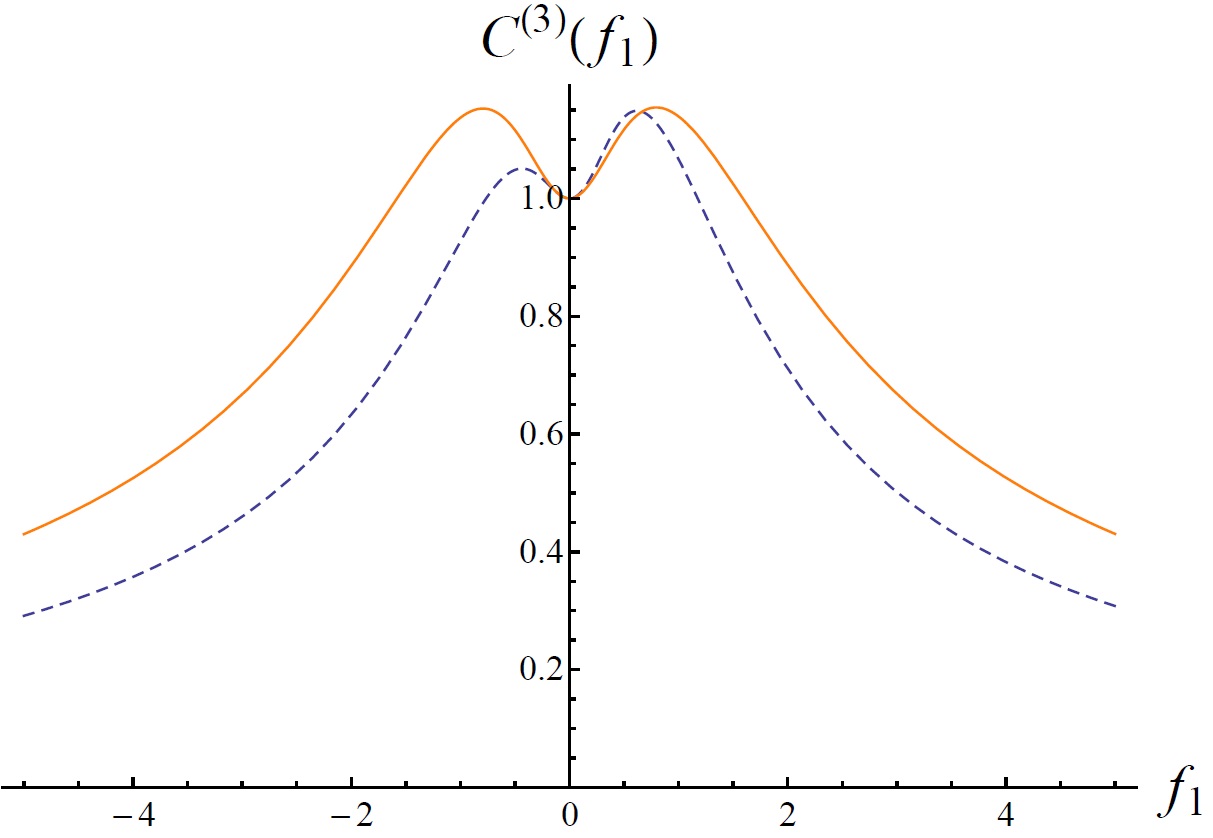}}
\caption{\small {(Color online) Concurrence of $|\Psi^{(3)}\rangle_{s}$ (dashed line) and $|\Psi^{(3)}\rangle_{c}$ (full line) as a function of $f_{1}$ for $\xi=6$, $\eta=2.5$, $\tau=5$ and $f_2=-1$ .} \label{qut} }
\end{figure}

\begin{figure}[ht]
\centerline{\includegraphics[width=16cm]{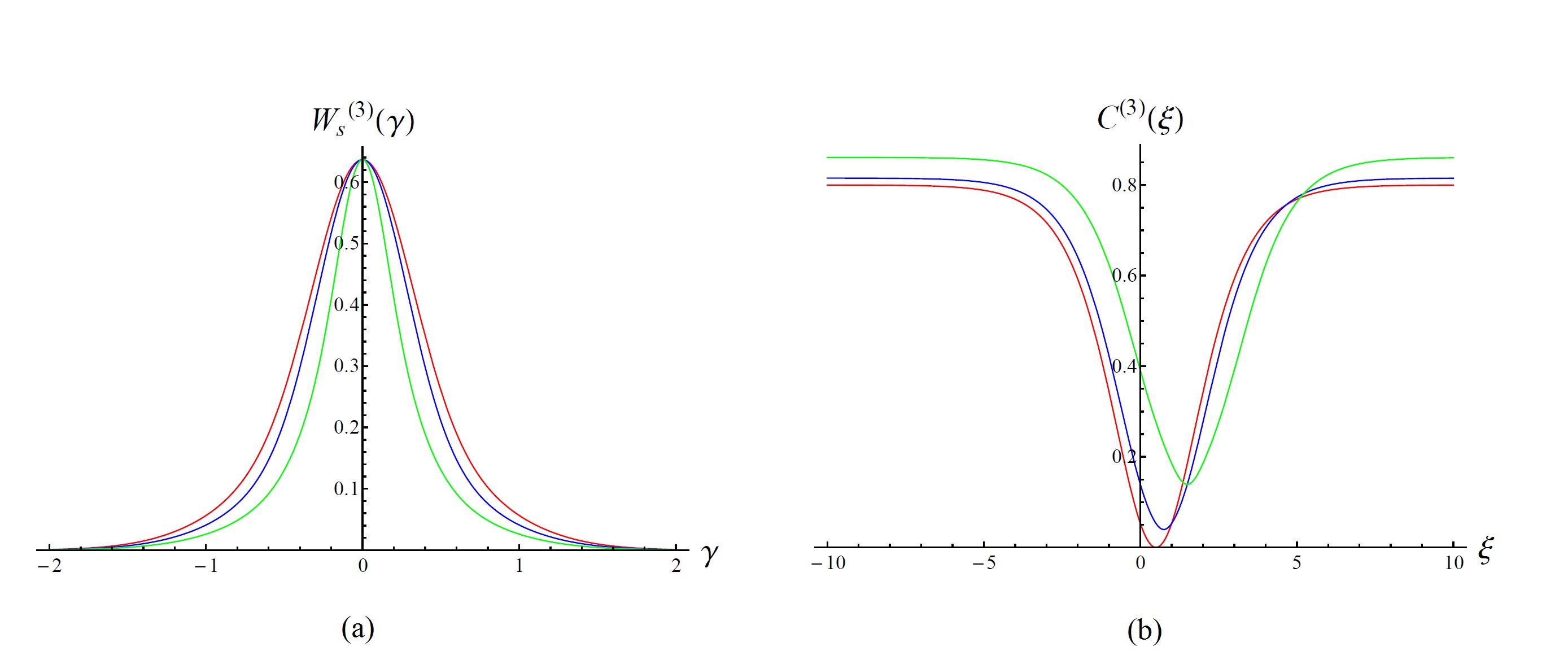}}
\caption{\small {(Color online) (a) Wigner function $W^{(3)}_s(\gamma)$ as a function of $\gamma$ with  $\xi=0.5$, (b) concurrence $C^{(3)}(\xi)$ as a function of $\xi$  with $f_1=f_2=1$ for given  $\eta=\tau=0.5$ (red), $\eta=0.5,\tau=1$ (blue), $\eta=1,\tau=2$ (green).} \label{8} }
\end{figure}

\begin{figure}[ht]
\centerline{\includegraphics[width=10cm]{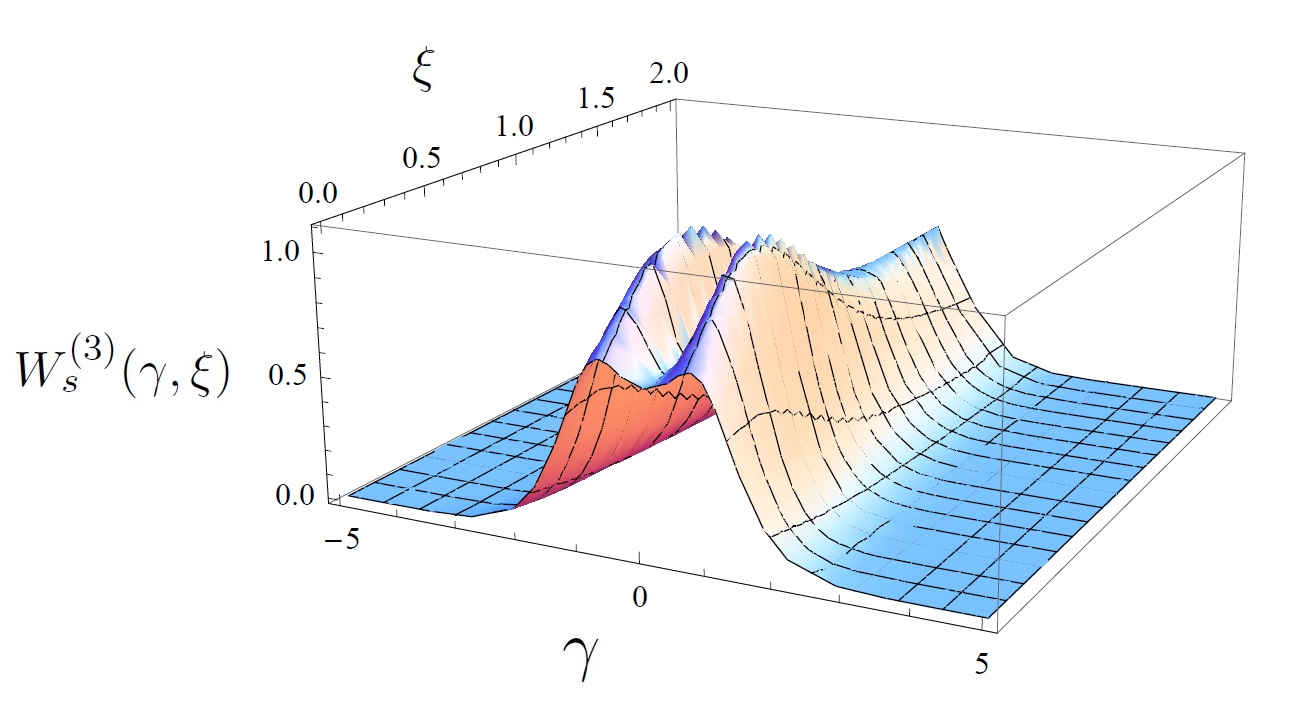}}
\caption{\small {(Color online) $W^{(3)}_s(\gamma,\xi)$ as a function of $\gamma$ and $\xi$ with $f_1=f_2=-1$ for given $\eta=0.2$ and $\tau=0.4$.} \label{9} }
\end{figure}

In this section, we consider two and three-mode qutrit like ESS's. First consider two-mode case:
\be\label{2}
|\Psi^{(3)}\rangle_{s}=\frac{1}{\sqrt{M_s^{(3)}}}(|\xi\xi\rangle+f_1|\eta\eta\rangle+f_2|\tau\tau\rangle),
\ee
where $M_s^{(3)}=1+f_1^2+f_2^2+2f_1p_1^2+2f_2p_3^2+2f_1f_2p_2^2$ is the normalization factor with $p_1=\langle\xi|\eta\rangle$, $p_2=\langle\tau|\eta\rangle$ and $p_3=\langle\tau|\xi\rangle$. For simplicity we assumed that all parameters are real.
In general, three squeezed states $|\xi\rangle$, $|\eta\rangle$ and $|\tau\rangle$,  can define three-dimensional space spanned by the set $\{|0\rangle,|1\rangle,|2\rangle\}$
\be\label{newbase}
\begin{array}{l}
|0\rangle=|\xi\rangle,\\
|1\rangle=\frac{1}{\sqrt{1-p_{1}^{2}}}(|\eta\rangle-p_{1}|\xi\rangle),\\
|2\rangle=\sqrt{\frac{1-p_{1}^{2}}{1-p^2_{1}-p^2_{2}-p^2_{3}+2p_1p_2p_3}}\left(|\tau\rangle+(\frac{p_{1}p_{3}-p_{2}}{{1-p_{1}^{2}}})|\eta\rangle+(\frac{p_{1}p_{2}-p_{3}}{{1-p_{1}^{2}}})|\xi\rangle\right).
\end{array}
\ee
Substitution these new basis into Eq.(\ref{2}) the state $|\Psi^{(3)}\rangle_{s}$ is reduced to a qutrit like ESS. We use the general concurrence measure for bipartite state $|\psi\rangle=\sum_{i=1}^{d_{1}}\sum_{j=1}^{d_{2}}a_{ij}|e_{i}\otimes e_{j}\rangle$  \cite{Akhtarshenas}. The norm of concurrence vector is obtained  as $C = 2 (\sum\limits_{i < j}^{d_1 } {\sum\limits_{k < l}^{d_2 } {\left| {a_{ik}a_{jl}-a_{il}a_{jk} } \right|^2 }})^{1/2}$, where $d_{1}$ and $d_{2}$ are dimensions of first and second part respectively (in the present case $d_{1}=d_{2}=3$). The concurrence $|\Psi^{(3)}\rangle_{s}$ is represented in figure \ref{qut}.
To compare this result with ECS, we also plotted the concurrence of $|\Psi^{(3)}\rangle_{c}$ in the figure \ref{qut}. By $|\Psi^{(3)}\rangle_{c}$ we mean that in  Eq.(\ref{2}) the squeezed states $|\xi\rangle,|\eta\rangle$ and $|\tau\rangle$ must be replaced by coherent states $|\alpha\rangle,|\beta\rangle$ and $|\gamma\rangle$.
Figure \ref{8}(a) displays the Wigner function of $|\Psi^{(3)}\rangle_{s}$ as a function of $\gamma$ with $f_1=f_2=1$ for fixed $\xi=0.5$ but  various $\eta$ and $\tau$. One may immediately deduce that  the the width of Wigner function decreases by increasing the distance of $\xi$ with $\eta$ and $\tau$. This result is confirmed by the concurrence measure  in a opposite manner which  is accompanied by a shift in $C^{(3)}(\xi)$ (see figure \ref{8}(b)).
Figure \ref{9} illustrates the $W^{(3)}_s(\gamma,\xi)$ for given  $\eta=0.2,\tau=0.4$ and  $f_1=f_2=-1$.
\subsection{Monogamy Inequality for Multi-Qutrit like ESS}
\begin{figure}[ht]
\centerline{\includegraphics[width=13cm]{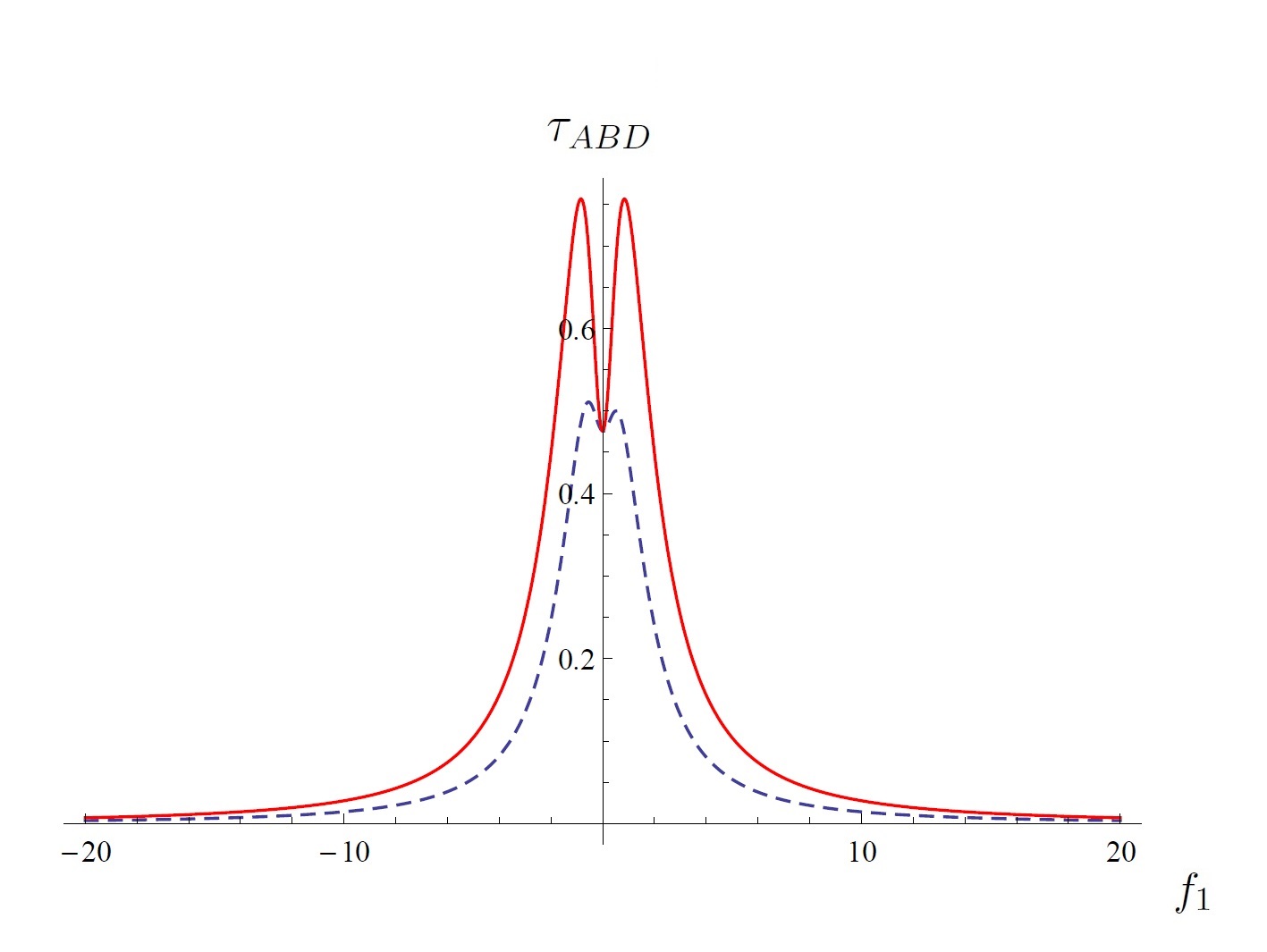}}
\caption{\small {(Color online) $\tau_{ABD}^{(s)}$ for qutrit like ESS (dashed line) and $\tau_{ABD}^{(c)}$ for qutrit like  ECS (full line) as a function of $f_1$ for given $N=20, m_1=1, m_2=2$ and $f_2=0.4$ for $\xi=3$, $\eta=2$.} \label{3} }
\end{figure}

Another problem that arises in multipartite states is monogamy inequality. Here in this section, we examine the concurrence based monogamy inequality for qutrit like ESS. To this end consider multi qutrit like ESS:
\be
|\Phi^{(3)}\rangle_{s}=\frac{1}{\sqrt{M'^{(3)}_N}}(|\xi\rangle...|\xi\rangle+f_1|\eta\rangle...|\eta\rangle+f_2|\tau\rangle...|\tau\rangle),
\ee
where
\be
M^{(3)}_N=1+f^{2}_1+f^{2}_2+2f_{1}p^{N}_1+2f_{2}p^{N}_3+2f_{1}f_{2}p^{N}_2,
\ee
is normalization factor. Once again,  we assume that the all parameters are real. We next  assume that the state $|\Phi^{(3)}\rangle_{s}$ as tripartite $A$, $B$ and $D$ including $m_1$, $m_2$ and $m_3=N-m_1-m_2$ modes respectively. For the moment, suppose that $\tau\rightarrow\infty$, i.e. $p_2,p_3=0$ and $p_1\neq0$. One  can easily obtain the reduced density matrices $\rho_{AB}=Tr_{_{D}}(|\Phi^{(3)}_{N}\rangle_{ABD}\langle\Phi^{(3)}_{N}|)$ and $\rho_{AD}=Tr_{_{B}}(|\Phi^{(3)}_{N}\rangle_{ABD}\langle\Phi^{(3)}_{N}|)$ by partially tracing out subsystems $D$ and $B$ respectively.  For general mixed states the concurrence is defined as $|C|^2=\sum_{\alpha\beta}|C_{\alpha\beta}|^2$ where $C^{\alpha\beta}=\lambda^{\alpha\beta}_1-\sum^n_{i=2}\lambda^{\alpha\beta}_i$ with $\lambda_1=\max\{\lambda_i, i=1,...,d^2\}$ and $\lambda^{\alpha\beta}_i$ are the nonnegative eigenvalues of $\tau\tau^*$ defined as \cite{YOU}
\be\label{yo}
\tau^{\alpha\beta}\tau^{\alpha\beta*}=\sqrt{\rho}(E_\alpha-E_{-\alpha})
\otimes(E_\beta-E_{-\beta})\rho^*(E_\alpha-E_{-\alpha})\otimes(E_\beta-E_{-\beta})\sqrt{\rho},
\ee
in which $\alpha$'s are positive roots of the lie group SU(N) (here for qutrit like  $SU(3))$ and $E_{\pm\alpha}$'s are corresponding   raising/lowering operators  (like $J_{\pm}$ of the angular momentum operator).
Let us consider $N=20, m_1=1, m_2=2$ and $f_2=0.4$ and explore the monogamy inequality \cite{coffman,kim,sanders1}
\be
\label{monogamy}
C_{A(BD)}^{2}\geq C_{AB}^{2}+C_{AD}^{2},
\ee
where $C_{AB}$ and $C_{AD}$ are the concurrences of the reduced density matrices  $\rho_{AB}$ and $\rho_{AD}$ respectively and
$C_{A(BD)}$ is the concurrence of pure state $|\Phi^{(3)}\rangle_{s}$ with respect to the partitions  $A$ and $BD$. The concurrence vectors $C_{AB}$ and $C_{AD}$ read
\be
\begin{array}{l}
C_{AB}=\frac{{{f_1}{p^{17}}\sqrt {(1 - {p^2})(1 - {p^4})} }}{{0.58 + 0.5f_1^2 + {f_1}{p^{20}}}},\\
C_{AD}=\frac{{{f_1}{p^2}\sqrt {(1 - {p^2})(1 - {p^{34}})} }}{{0.58 + 0.5f_1^2 + {f_1}{p^{20}}}},
\end{array}
\ee
where $p=\frac{1}{\sqrt{\cosh r_1\cosh r_2(1-\tanh r_1\tanh r_2)}}$. On the other hand, one finds that
\be
C_{A(BD)}=\frac{{\sqrt {0.64 + 1.28{f _1}{p^{20}} + f _1^2(4.64 - 4{p^2} - 4{p^{38}} + 4{p^{40}})} }}{{1.16 + f _1^2 + 2{f _1}{p^{20}}}}.
\ee
The behavior of $\tau_{ABD}=C_{A(BD)}^{2}-C^2_{AB}-C^2_{AD}$ as a function of $f_1$, for given $\xi$ and $\eta$, is shown in figure \ref{3}.
In order to compare this results with ECS we also display the results of $\tau_{ABD}^{(c)}$ calculated in Ref. \cite{najarbashi1}.
Figure \ref{3} shows that in general   $\tau_{ABD}^{(c)}>\tau_{ABD}^{(s)}$.
\par
\section{Conclusion}
In summary, we introduced two-mode qubit and qutrit like ESS's. For qubit like state $|\Psi^{(2)}\rangle_s$, a comparison between ESS and ECS showed that for $f>0$ the concurrence of squeezed state is smaller than the concurrence of coherent state without crossing. On the other hand for $f<0$, there are  crossing for some parameters $r_1$, $r_2$ and $\Delta\theta$. It was shown  the concurrence is a monotone function of $\Delta_s$ for  all real parameters  and $f=1$ with the properties that when $\Delta_s\rightarrow\infty$, the concurrence tends to its maximum value ($C^{(2)}_{max}=1$), while for
small separation (i.e. $\Delta_s\rightarrow0$) the concurrence tends to zero and the state becomes separable.
 Moreover we investigated the ESS generated   by beam splitters, phase-shifter and cross Kerr nonlinearity and showed that for $r_1=r_2\neq0$ (i.e. $p_1=p_2$) the state $|\psi^{(2)}\rangle_{s}$ is maximally entangled state (i.e. $C^{(2)}_{max}=1$).
Comparing Wigner functions   $W_s^{(2)}(\gamma)$ and $W_c^{(2)}(\gamma)$ revealed that  for $f=1$ and $f=-1$, unlike the $W_c^{(2)}(\gamma)$, the peaks of $W_s^{(2)}(\gamma)$ approached to each other by increasing the difference of $\Delta_s=\xi-\eta$. This behavior is different for $W_c^{(2)}(\gamma)$, i.e. the peaks of $W_c^{(2)}(\gamma)$ recede from each other by increasing the $\Delta_c=\alpha-\beta$. Similar results hold for qutrit like ESS, $|\Psi^{(3)}\rangle_s$. 
Comparing the the concurrences of ESS to that of the ECS, showed that for $f_2=-1$ there are $\xi,\eta$ and $\tau$  which cross over occurs. Furthermore, for $f_1=f_2=1$, it was shown  that by increasing the distances of $\xi,\eta$ and $\tau$, the width of $W^{(3)}_s(\gamma)$ decreases whereas the concurrence $C^{(3)}(\xi)$ has upper-shift. Finally, we explored  the  monogamy inequality for both qutrit like ESS and ECS. Although in both cases the monogamy inequality was not violated, but in general the inequality $\tau_{ABD}^{(c)}>\tau_{ABD}^{(s)}$ is satisfied.

\par
\textbf{Acknowledgments}\\
The authors also acknowledge the support from the University of Mohaghegh Ardabili.

\end{document}